\documentclass{aa}
\usepackage{graphicx,epsfig,fancyhdr,psfig,rotating,amsmath,epsf,txfonts,natbib,epstopdf,multirow,booktabs,longtable,color,soul}
\usepackage[switch]{lineno}

\def \kms {{\rm km\;s$^{-1}$}}

\def \arcsec {$^{''}$}

\def \al {$\alpha$}

\graphicspath{{./}}

\begin{document}


\title{Numerous Bidirectionally Propagating Plasma Blobs near the Reconnection Site of a Solar Eruption}

\author{Zhenyong Hou\inst{1}
          \and
          Hui Tian\inst{1}
          \and
          Maria S. Madjarska\inst{2,3}
          \and
          Hechao Chen\inst{4}
          \and
          Tanmoy Samanta\inst{5}
          \and
          Xianyong Bai\inst{6,7}
           \and
          Zhentong Li\inst{8}
          \and
          Yang Su\inst{8}
          \and
          Wei Chen\inst{8}
          \and
          Yuanyong Deng\inst{6,7}
}

\institute{School of Earth and Space Sciences, Peking University, Beijing, 100871, China\\ \email{huitian@pku.edu.cn}
         \and
         Max Planck Institute for Solar System Research, Justus-von-Liebig-Weg 3, 37077, G\"ottingen, Germany
         \and
         Space Research and Technology Institute, Bulgarian Academy of Sciences, Acad. Georgy Bonchev Str., Bl. 1, 1113, Sofia, Bulgaria
         \and
         School of Physics and Astronomy, Yunnan University, Kunming 650050, China
         \and 
         Indian Institute of Astrophysics, Koramangala II Block, Bangalore-560034, India
         \and
         National Astronomical Observatories, Chinese Academy of Sciences, Beijing, 100011, China
         \and 
         School of Astronomy and Space Science, University of Chinese Academy of Sciences, Beijing 101408, China
         \and 
         Key Laboratory of Dark Matter and Space Astronomy, Purple Mountain Observatory, Chinese Academy of Sciences (CAS), Nanjing 210023, People's Republic of China
}

\date{Received date, accepted date}

\abstract
{Current sheet is a common structure involved in solar eruptions.
However, it is observed in minority of the events and the physical properties of its fine structures during a solar eruption are rarely investigated.
Here, we report an on-disk observation that displays 108 compact, circular or elliptic bright structures, presumably plasma blobs, propagating bidirectionally along a flare current sheet during a period of $\sim$24 minutes.
From extreme ultraviolet images, we have investigated the temporal variation of the blob number around the flare peak time.
The current sheet connects the flare loops and the erupting filament.
The width, duration, projected velocity, temperature, and density of these blobs are $\sim$1.7$\pm$0.5\,Mm, $\sim$79$\pm$57\,s, $\sim$191$\pm$81\,\kms, $\sim$10$^{6.4\pm0.1}$ K, and $\sim$10$^{10.1\pm0.3}$ cm$^{-3}$, respectively.
The reconnection site rises with a velocity of $\leqslant$69\,\kms.
The observational results suggest that plasmoid instability plays an important role in the energy release process of solar eruptions.}

\authorrunning{Hou et al.}
\titlerunning{Numerous Bidirectional propagating plasma blobs}

\maketitle

\section{Introduction}
\label{sec:intro}
Magnetic reconnection is a fundamental process that is involved in various solar activities 
  \citep[e.g.,][]{1991JGR....96.9399D,1997Natur.386..811I,2007Sci...318.1591S,2014Sci...346A.315T,2016NatPh..12..847L,2016NatCo...711837X,2017ApJ...851...67S,2018ApJ...864L...4L,2018ApJ...854..174T,2018ApJ...854...92T,2019STP.....5b..58H,2019ApJ...887..137S,2021ApJ...915...39H,2021ApJ...918L..20H,2022A&A...660A..45M,2023NatCo..14.2107C,2023ApJ...958..116W}.
When a large-scale solar eruption occurs, the closed magnetic field is severely stretched out by the ejected structure.
As a result, a current sheet generally forms, that is a narrow region across which the magnetic field changes rapidly
  \citep[see][]{2000JGR...105.2375L,2002A&ARv..10..313P,2003NewAR..47...53L,2006SSRv..123..251F}.
Magnetic reconnection can cause the magnetic field reconfiguration and the release of magnetic energy stored in the system beforehand.
Many observational features related to the reconnection process have been observed \citep{2015SSRv..194..237L}, e.g., 
  current sheets
  \citep[e.g.,][]{2002ApJ...575.1116C,2011ApJ...727L..52R,2013ApJ...767..168L,2017ApJ...835..139S,2018ApJ...854..122W,2021ApJ...911..133C,2024ApJ...964...58D}, 
  inflows \citep[e.g.,][]{2001ApJ...546L..69Y,2005ApJ...622.1251L,2006ApJ...637.1122N,2009ApJ...703..877L,2012ApJ...754...13S},
  sunward outflows \citep[e.g.,][]{2014ApJ...797L..14T,2015NatCo...6.7598S},
  anti-sunward outflows \citep[e.g.,][]{2007ApJ...661L.207W,2017ApJ...841...49C},
  and bidirectional outflows
  \citep[e.g.,][]{2010ApJ...722..329S,2013NatPh...9..489S,2016ApJ...821L..29Z,2018ApJ...853L..18Y}.
The plasma that surrounds the current sheets can be heated to temperatures of several or tens of millions of Kelvin
  \citep[e.g.,][]{2013NatPh...9..489S,2018ApJ...853L..15L,2023SoPh..298...61R}.
The current sheets of solar eruptions may also experience transverse oscillations \citep[e.g.,][]{2016ApJ...829L..33L,2022ApJ...931L..32W}.

Theoretical studies and magnetohydrodynamic simulations have suggested that 
  the onset of plasmoid instability plays a crucial role in fast energy releasing in current sheets
  \citep[e.g.,][]{2009PhPl...16k2102B,2012ApJ...760...81K,2015ApJ...799...79N,2019MNRAS.482..588Y,2022RAA....22h5010Z}.
Plasma blobs (blob-like hot plasma) are believed to be an observable proxy of the occurrence of plasmoid instability in current sheets of solar eruptions.
However, imaging fine structures within the current sheets of solar eruptions is very hard, especially using original intensity images.
White-light observations have shown some anti-sunward propagating blobs appearing along current sheets
  \citep[e.g.,][]{2003ApJ...594.1068K,2005ApJ...622.1251L,2010ApJ...723L..28L,2012SoPh..276..261S,2013ApJ...771L..14G,2014ApJ...784...91L,2016ApJ...826...94K,2016ApJ...831...47S,2016SoPh..291.3725W,2020ApJ...892..129L}.
The white-light blobs usually are far away from the reconnection site and appear as diffuse structures with a width of $\geqslant$10~Mm in post coronal mass ejection (CME) current sheets.
After the launch of the Solar Dynamics Observatory \citep[SDO,][]{2012SoPh..275....3P},
  the extreme ultraviolet (EUV) images taken by the Atmospheric Imaging Assembly
  \citep[AIA,][]{2012SoPh..275...17L} revealed sporadic small-scale plasma blobs propagating along current sheets of solar eruptions
\citep{2012ApJ...745L...6T,2013A&A...557A.115K,2013MNRAS.434.1309L,2016ApJ...821L..29Z,2019SciA....5.7004G,2018ApJ...869..118D,2022ApJ...924L...7L,2023ApJ...943..156K}.
\cite{2013A&A...557A.115K} reported the first simultaneous EUV and radio observations of several bidirectionally moving plasma blobs in a solar flare.
Among recurring bidirectional outflows along the current sheet associated with a C2.0 flare,
  \cite{2016ApJ...821L..29Z} found only two sunward propagating plasma blobs.
\cite{2019SciA....5.7004G}  found that anti-sunward moving plasma blobs merged into a larger blob that later evolved into a CME bubble.
With the combination of EUV and white-light observations, \cite{2020A&A...644A.158P} carried out a statistical research on
  the bidirectionally propagating blobs during the post-impulsive phase of a long-duration solar eruption \citep{2020ApJ...900...17Y}.
They identified only 20 sunward plasma blobs and 16 anti-sunward ones in the enhanced EUV intensity images during a period of $\sim$2 hours.
Recently, \cite{2023ApJ...943..156K} reported a series of bidirectionally propagating plasmoids continuously released for about 30 minutes in a vertical flare current sheet behind an erupting flux rope.
However, characterizing the role of plasmoid instability in the energy release of magnetic reconnection of solar eruptions is still a challenge.

In this study, we analyze multi-passband EUV images of a current sheet during an eruptive solar flare.
Around the flare peak time, we identify more than one hundred compact,
  bidirectionally propagating plasma blobs along the current sheet, investigate their physical properties, and track their subsequent temporal evolution.
We describe the observations in Section\,\ref{sec:obs}, present the analysis results in Section\,\ref{sec:res}, and the discussion in Section\,\ref{sec:dis}.
We summarize our findings in Section\,\ref{sec:sum}.

\section{Observations}
\label{sec:obs}

On November 19, 2022, a solar eruption was simultaneously detected by AIA on board SDO,
  the Solar Upper Transition Region Imager \citep[SUTRI,][]{2023RAA....23f5014B,2023RAA....23i5009W}
  on board the Space Advanced Technology demonstration satellite (SATech-01), the Hard X-ray Imager
  \citep[HXI,][]{2019RAA....19..163S,2019RAA....19..160Z} on board the Advanced Space-Based Solar Observatory
  \citep[ASO-S,][]{2019RAA....19..156G,2023SoPh..298...68G}, and the Geostationary Operational Environmental Satellite (GOES).
We mainly used the EUV images taken by AIA and SUTRI, to analyze the dynamics of the current sheet of the solar eruption.
The AIA EUV images with different response temperatures have a pixel size of 0.6\arcsec\ and a cadence of 12~s.
The SUTRI 465\,\AA\ images capture the full-disk transition region at a temperature regime of $\sim$0.5 MK \citep{2017RAA....17..110T}
  and have a pixel size of 1.2\arcsec\ and a cadence of 30 s.
The SUTRI 465 \,\AA\ and AIA 304 \,\AA\ images were aligned using a linear Pearson correlation analysis.
The AIA 1600\,\AA\ passband with a pixel size of 0.6\arcsec\ and a cadence of 24 s, is also used to identify the flare ribbons of the solar eruption.

For this study, we reconstructed HXI image in the energy range of 20--30 keV with a pixel size of 2\arcsec\ using a HXI Clean algorithm and the subcollimator groups G3 to G10.
The HXI image provides a spatial resolution of $\sim$6.5\arcsec.
The data observed 48\,h after the imaging time is taken as background.
Only the HXI 20--30 keV emisstion near 12:53 UT was reconstructed with a 20-second integration time,
  while the rest of the data did not have reliable quality.
The images between HXI and AIA were aligned by using reference structures, such as flare ribbons and brightenings above flare loops.

For the identification of an associated CME of the solar eruption, we used the coronagraph images taken by the Large Angle Spectroscopic Coronagraph
  \citep{1995SoPh..162..357B} on board the Solar and Heliospheric Observatory \citep[SOHO,][]{1998GeoRL..25.2465T}.

\section{Results}
\label{sec:res}

\subsection{Bidirectional flows along the current sheet}
\label{subsec:flow}

\begin{figure*}
\centering
\includegraphics[trim=0.0cm 0.3cm 0.0cm 0.0cm,width=1.0\textwidth]{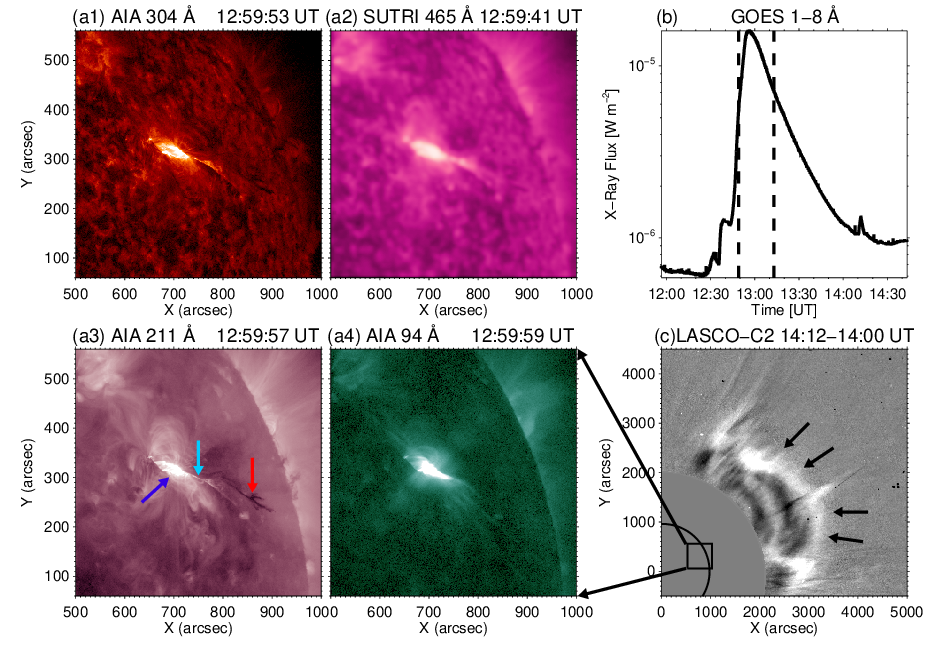}
\caption{An overview of the solar eruption.
(a1)--(a4) AIA 304\,\AA, SUTRI 465\,\AA, AIA 211\,\AA, and 94\,\AA\ images at $\sim$13:00:00 UT.
In (a3), the blue, cyan, and red arrows mark the flare loops, current sheet, and rising filament, respectively.
(b) Light curve of GOES X-ray flux at 1--8\,\AA.
The vertical dashed lines indicate the time interval when the current sheet is visible in the EUV images.
(c) The LASCO C2 difference image between 14:12 UT and 14:00 UT showing the associated CME as marked by the black arrows.
The black box indicates the field of view in (a1)--(a4) and the black curve represents the solar limb.}
\label{fig:overview}
\end{figure*}

\begin{figure*}
\centering
\includegraphics[trim=0.0cm 0.3cm 0.0cm 0.0cm,width=0.9\textwidth]{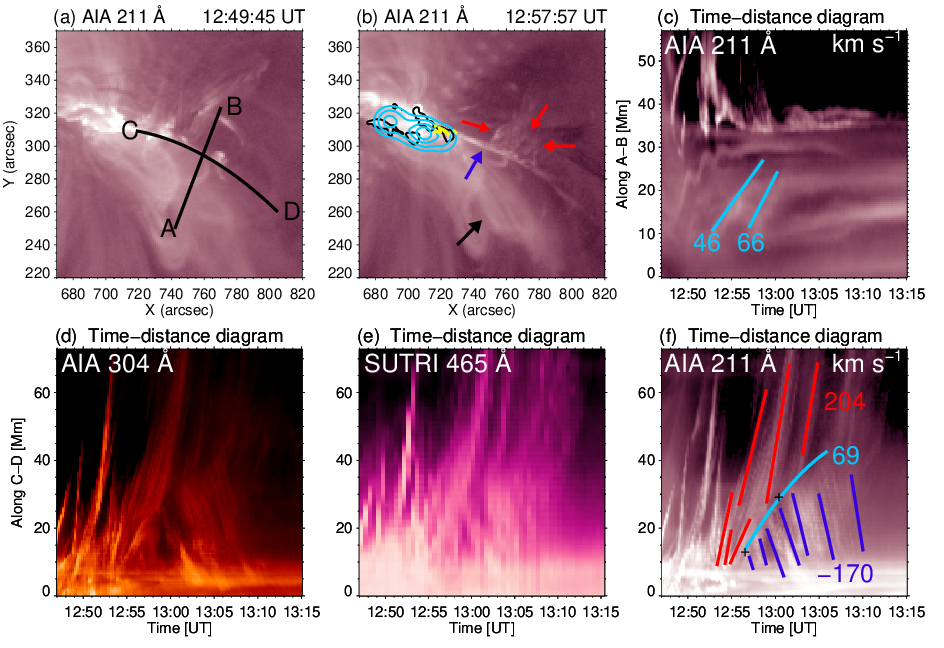}
\caption{Plasma flows observed during the solar eruption.
(a) and (b) AIA 211\,\AA\ images at 12:49:45~UT and 12:57:57~UT, respectively.
In (a), cut A--B is used to construct the time-distance diagram in (c), while cut C--D is used to construct the time-distance diagrams in (d), (e), and (f).
In (b), the black arrow indicates coronal loops which are observed continuously moving to the current sheet during the solar eruption,
  and the blue and red arrows indicate the main current sheet and splited ones, respectively.
The black and cyan contours in (b) indicate the enhanced AIA 1600\,\AA\ emission and the HXI 20--30 keV source near 12:53 UT, respectively.
The yellow dot-like symbols represent some of the sunward moving blobs.
The cyan curves in (c) indicate the motions of the coronal loops.
In (f), the cyan curve marks the rising motion of the reconnection site, while the blue and red lines indicate the bidirectional flows in the current sheet.
The two black plus symbols mark the section of the curve that is used to calculate the maximum velocity.
The numbers in (c) and (f) represent the velocities of the flows  in the unit of \kms.}
\label{fig:flows}
\end{figure*}

\begin{figure*}
\centering
\includegraphics[trim=0.0cm 0.0cm 0.0cm 0.0cm,width=0.9\textwidth]{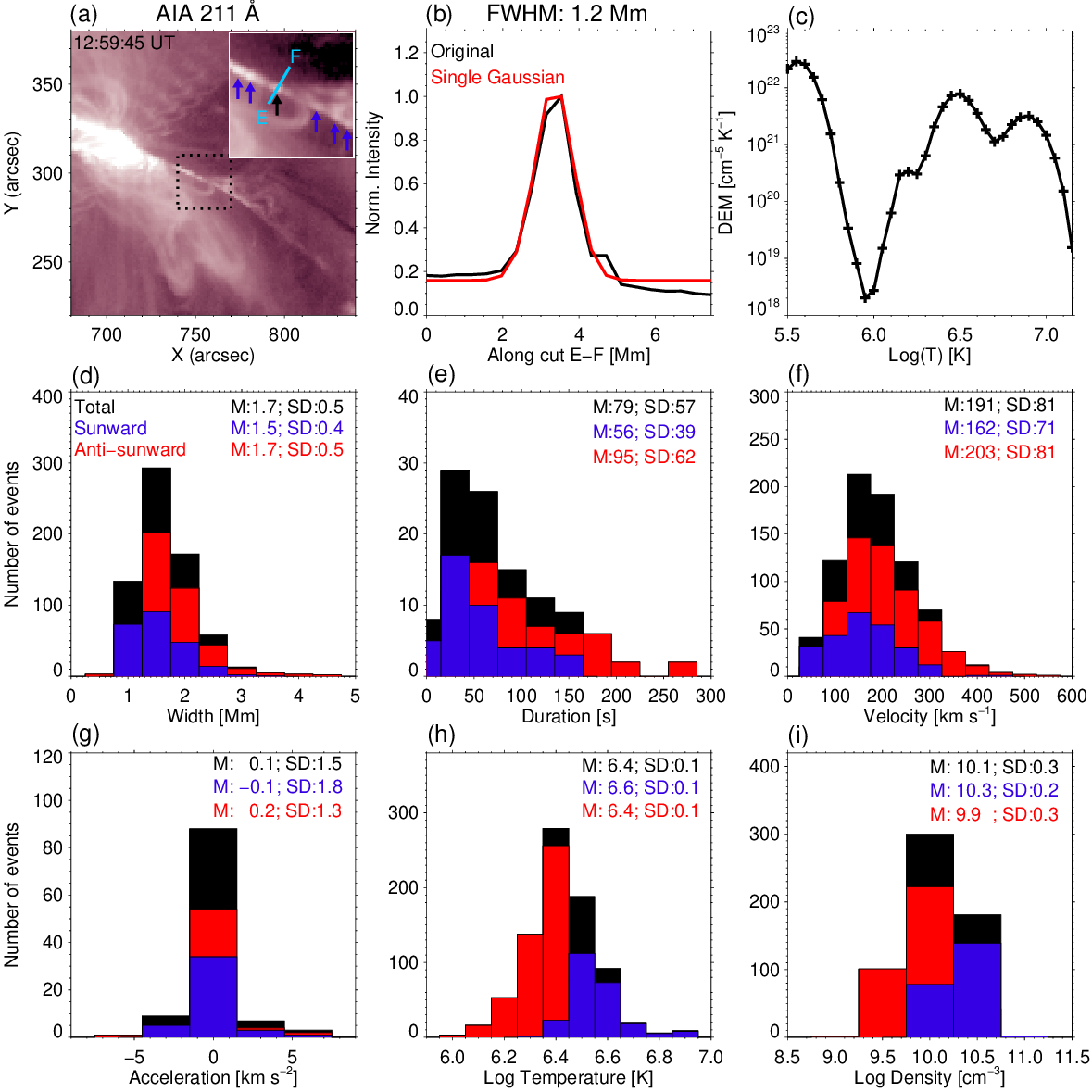}
\caption{Plasma blobs identified in the current sheet and their parameters.
(a) AIA 211\,\AA\ image at 12:59:45 UT showing several plasma blobs.
The region of the inset in (a) is marked by the black box.
The blue and black arrows in the inset indicate several plasma blobs.
(b) Normalized intensity variation (black) along cut E--F shown in (a) and its Gaussian fit (red).
(c) DEM distribution of the blob marked by the black arrow in (a).
(d)--(i) Distributions of the plasma blob parameters.
The black histogram is obtained from the total plasma blobs,
  while the blue and red ones are obtained from the sunward and anti-sunward moving blobs, respectively.
`M' and `SD' represent the average values and standard deviations, respectively.
An animation of the AIA 211 \AA\ images and the ones processed by MGN method are available, showing the bidirectionally propagating plasma blobs.
It covers a duration of $\sim$24 minutes from 12:48:57 UT to 13:13:33 UT.
In the left and middle panels of the animation, the dot-like symbols mark the blob locations, while the blue and green numbers represent the sunward and anti-sunward moving blobs, respectively.}
\label{fig:blobs_paras}
\end{figure*}


Figure\,\ref{fig:overview} presents an overview of the solar eruption that appears as a filament eruption in the EUV images taken by AIA and SUTRI.
An M1.6 flare was detected by GOES, as shown in Figure\,\ref{fig:overview}(b).
One hour later, an associated CME is visible in the white-light coronagraph image, indicated by the black arrows in Figure\,\ref{fig:overview}(c).

Around the peak time of this eruptive flare, from 12:49~UT to 13:13~UT marked by the dashed lines in Figure\,\ref{fig:overview}(b),
  a sheet-like structure with a projected length of $\leqslant$80~Mm appears in the EUV images and
  connects the bright flare loops and the rising filament (Figure\,\ref{fig:overview}(a1)--(a4)).
During this time interval, we can also see many bidirectionally propagating plasma blobs along this sheet-like structure in the EUV passbands (see Section\,\ref{subsec:blob}).
This sheet-like structure is likely consistent with a current sheet in the low corona.
The signature of this current sheet is prominently visible in AIA 304\,\AA\ (10$^{4.7}$ K), SUTRI 465\,\AA\ (10$^{5.7}$ K), and AIA 211\,\AA\ (10$^{6.3}$ K),
  indicating that the current sheet has a multi-thermal structure.
The AIA 211\,\AA\ images reveal the current sheet very well, allowing us to investigate its dynamics and properties.
The current sheet is weakly visible in the high-temperature passbands, such as AIA 131\,\AA\ (10$^{5.6}$ K and 10$^{7.0}$ K) and 94\,\AA\ (10$^{6.8}$ K),
  implying that the current sheet contains a small amount of plasma at high temperature.

Figure\,\ref{fig:flows} shows the recurring, bidirectional plasma flows along the current sheet
  around the peak time of the eruptive flare in the EUV passbands of SUTRI and AIA.
These flows are the apparent motions of plasma blobs that are likely the results of magnetic reconnection occurring within the current sheet.
From 12:52 UT to 13:01 UT, a group of coronal loops is apparently moving towards the current sheet (see the associated animation of Figure\,\ref{fig:blobs_paras}),
  which is indicated by a black arrow in Figure\,\ref{fig:flows}(b).
The approaching motion of the coronal loops toward the current sheet may be a signature of inflows toward the reconnection site.
Around 12:54 UT, several fibril-like structures begin to split from the upper part of the current sheet, as indicated by the red arrows in Figure\,\ref{fig:flows}(b).
These fine structures involve many anti-sunward flows,
which may resemble coronal nanojets that are characterized as transient unidirectional jet-like features perpendicular to the coronal loop \citep[e.g.,][]{2020ApJ...899...19C,2021NatAs...5...54A}.

The HXI observation reveals two hard X-ray sources in the HXI 20--30 keV range (cyan contour lines in Figure\,\ref{fig:flows}(b)).
In Figure\,\ref{fig:flows}(b), we also plotted the AIA 1600\,\AA\ contours as the black lines to show the flare ribbons and some sunward moving blobs as the yellow dot-like symbols.
The western HXI source appears to cover the flare loops and the sunward outflow region within the current sheet,
  indicating a possible Masuda-type coronal source \citep[e.g.,][]{1994Natur.371..495M} and a source in the reconnection outflow region,
  possibly produced by electrons confined in the plasma blob \citep{2022ApJ...933...93K}.
It may also indicate multiple sources of energetic electrons in the loop top region due to the interactions of 
  plasma blobs with the flare termination shocks, as simulated in \cite{2020ApJ...905L..16K}.

To quantitatively determine the properties of the flows, we defined two cuts in Figure\,\ref{fig:flows}(a).
Cut A--B is along the direction of the motion of the coronal loops, and cut C--D is almost parallel to the current sheet.
Using these two cuts, we constructed the time-distance diagrams for the images of AIA 304\,\AA, SUTRI 465\,\AA, and AIA 211\,\AA\ passbands, as shown in Figure\,\ref{fig:flows}(c)--(d).

The time-distance diagram in Figure\,\ref{fig:flows}(c) illustrates the tracks of the moving coronal loops toward the current sheet.
Two cyan lines are used to mark the propagation tracks that may be the apparent signature of the inflows toward the reconnection site.
The velocities of the loop motions can be determined by applying linear fits to the propagation tracks,
  and are estimated to be 46--66\,\kms\ with an average value of $\sim$56\,\kms.

Figure\,\ref{fig:flows}(d)--(f) display the propagation of the bidirectioinal flows, i.e., the sunward flows and anti-sunward ones.
It can be seen that the bidirectional flows are similar in AIA 304\,\AA, SUTRI 465\,\AA, and AIA 211\,\AA.
We further identified several bidirectional propagation tracks to determine their velocities, as indicated by the blue and red lines in Figure\,\ref{fig:flows}(f).
By applying linear fits to the propagation tracks, the average velocities of the sunward and anti-sunward flows (i.e., outflow velocities) are estimated to be -170\,\kms\ and 204\,\kms, respectively.
The reconnection rate can be estimated as the velocity ratio of the inflows and the outflows.
Considering that we estimate the inflow only on one side of the reconnection site, the inflow velocity is estimated at 28\,\kms.
Therefore, the reconnection rate ranges from 0.14 to 0.16,
  which is consistent with the results inferred from off-limb eruptive observations
  \citep[e.g.,][]{2005ApJ...622.1251L,2013NatPh...9..489S},
  and larger than other results inferred from on-disk coronal or chromospheric observations
  \citep[e.g.,][]{2019ApJ...879...74C,2022NatCo..13..640Y}.
The calculation of the reconnection rates could be affected of the viewing angle, i.e. the projection effect.
The reconnection sites in \cite{2019ApJ...879...74C}, \cite{2022NatCo..13..640Y}, and our study also have different heights and extensions,
  which may to some extent affect the calculation of the projected velocities of the inflows and outflows.

Figure\,\ref{fig:flows}(d)--(f) also show another remarkable propagation track marked by the cyan line in Figure\,\ref{fig:flows}(f),
  representing the rise of the reconnection site that separates the sunward and the anti-sunward flows.
As derived from a linear fit to the steepest section of this track between the two plus symbols in Figure\,\ref{fig:flows}(f), the rising velocity of the reconnection site is estimated to be $\leqslant$69\,\kms.

\begin{figure*}
\centering
\includegraphics[trim=0.0cm 0.3cm 0.0cm 0.0cm,width=0.7\textwidth]{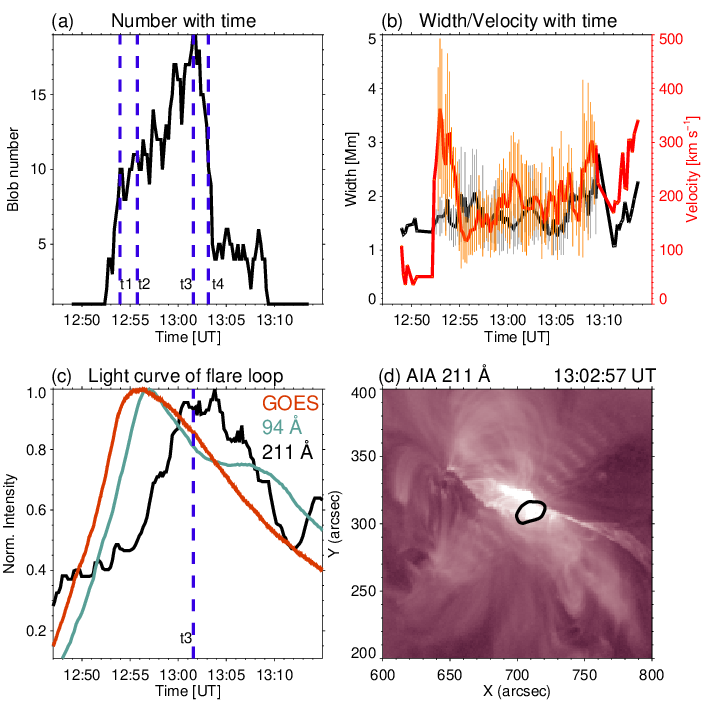}
\caption{Variations of the parameters for the plasma blobs.
(a) Variation of the blob number.
The blue dashed lines represent the four instances (t1, t2, t3, and t4) that are used for the power spectrum analysis in Figure\,\ref{fig:powerindex}.
(b) Variations of the blob width (black curve) and velocity (red curve).
In (b), the vertical grey and orange lines indicate the standard errors.
(c) Normalized intensity variations of GOES 1--8\,\AA\ and AIA 94/211 \AA\ in the region marked by the black contour in (d).
In (a) and (c), the blue dashed lines marked by `t3' also represent the peak instance of the blob number.
(d) AIA 211\,\AA\ image at 13:02:57 UT, showing a region (black contour) near the bottom of the current sheet.}
\label{fig:evolution}
\end{figure*}

\subsection{Bidirectionally propagating plasma blobs along the current sheet}
\label{subsec:blob}

\begin{figure*}
\centering
\includegraphics[trim=0.0cm 0.3cm 0.0cm 0.0cm,width=0.97\textwidth]{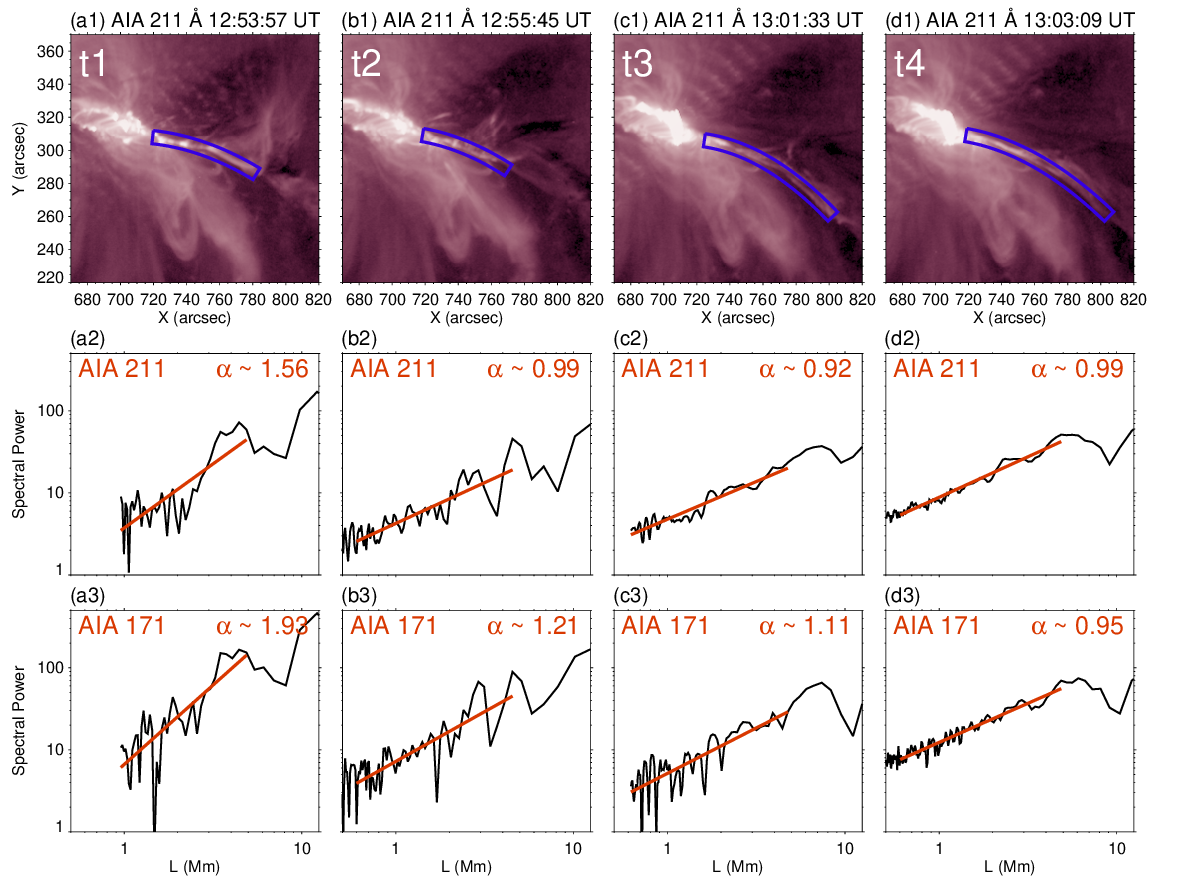}
\caption{Power spectrum analysis for the AIA 211\,\AA\ and 171\,\AA\ intensity along the current sheet.
(a1)--(d1) AIA 211\,\AA\ images at t1, t2, t3, and t4.
(a2)--(d2) Spectrum distributions of the AIA 211\,\AA\ intensity along the long side of the blue regions in (a1)--(d1) in the spatial frequency domain.
(a3)--(d3) Spectrum distributions of the AIA 171\,\AA\ intensity along the long side of the blue regions in (a1)--(d1) in the spatial frequency domain.
The red lines represent a fitting power-law behavior in the spatial range of 0.6--5 Mm.}
\label{fig:powerindex}
\end{figure*}

The occurrence of  magnetic reconnection is strongly supported by the visual evidence of fine structures in the current sheet,
  e.g., bidirectionally propagating blobs away from the reconnection site in the EUV images.
We manually identified 108 plasma blobs in the current sheet, i.e., 43 sunward blobs and 65 anti-sunward ones.
Several examples are shown in Figure\,\ref{fig:blobs_paras}(a), appearing as compact, circular/elliptic bright structures.
For all identified plasma blobs, we refer to the associated animation of Figure\,\ref{fig:blobs_paras},
  in which the blobs are marked by the dot-like symbols and located below the numbers (blue: sunward, green: anti-sunward).
We also applied the Multiscale Gaussian Normalization (MGN) method \citep{2014SoPh..289.2945M} to process the AIA 211\,\AA\ images.
The MGN method enhanced the appearance of these plasma blobs in the processed AIA 211\,\AA\ images (see the associated animation of Figure\,\ref{fig:blobs_paras}).
In the previous studies, propagating plasma blobs have been detected in coronal jets \citep[e.g.,][]{2012ApJ...759...33S,2014A&A...567A..11Z,2016SoPh..291..859Z,2022FrASS...8..238C,2024MNRAS.528.1094Y}.

From the AIA 211\,\AA\ images, we measured some of the physical parameters for these plasma blobs, i.e., the width, duration, projected velocity, acceleration, EM-weighted temperature, and density.
The parameter distributions are shown in Figure\,\ref{fig:blobs_paras}(c)--(h).
To estimate the width of a blob, we took a cut across the blob, e.g., cut E--F in Figure\,\ref{fig:blobs_paras}(a).
We then plotted the intensity profile along the cut as the black line in Figure\,\ref{fig:blobs_paras}(b).
After applying a single Gaussian fit to the intensity profile, we took the full width at half maximum (FWHM) as the width of the blob.
The duration of a blob is calculated as the time difference between its first appearance and its disappearance.
Due to the dynamic evolution of the current sheet, the start and end times of a blob may be not consistent with its identified appearance and disappearance times in the EUV images.
Thus, the obtained duration of a blob may be shorter than its lifetime.
The position of a blob is found to be the pixel with the maximum intensity, which was used to calculate the propagation velocity and acceleration of each blob.
If a blob exists in more than two AIA 211\,\AA\ frames, its acceleration can be calculated.
For the width, velocity, temperature, and density distributions, all the values of each blob at different observing times were considered.
On the other hand, there is only one duration and acceleration value for a propagating blob.
To examine the temperature and density of the blobs, we also performed a differential emission measure (DEM) analysis \citep{2015ApJ...807..143C,2018ApJ...856L..17S,2020ApJ...898...88X,2021Innov...200083S} for the current sheet region.
An example of a DEM distribution for one plasma blob is shown in Figure\,\ref{fig:blobs_paras}(c).

In Figure\,\ref{fig:blobs_paras}(c)--(h), the black-filled histograms are plotted for all the plasma blobs, while the blue and red ones are plotted for the sunward and anti-sunward propagating blobs, respectively.
The blob widths are mostly in the range of 0.8--3.0 Mm, with an average value of $\sim$1.7 Mm.
The smallest blob width of 0.6 Mm, which is close to the resolution limit of AIA,
  may be an indication that even smaller blobs exist but are unresolved in the current sheet.
The blob durations are shorter than 300~s, with an average value of 73\,s.
Their projected velocities are in the range of 50--550\,\kms\ and the average value is $\sim$190\,\kms.
Most blobs have little acceleration during their propagation.
These plasma blobs have a temperature of 10$^{6.0}$--10$^{6.9}$ K, with an average of 10$^{6.4}$ K.
Taking the width of each blob individually as the line-of-sight (LOS) integration length, the electron density was estimated to be from 10$^{9.4}$ to 10$^{11.6}$ cm$^{-3}$.
The distributions reveal clear differences between the sunward and anti-sunward blobs.
Compared to the anti-sunward blobs, the sunward ones experience a shorter propagating distance and a denser environment,
  which may lead to duration and velocity differences.

We also investigated the evolution of the blob parameters shown in Figure\,\ref{fig:evolution}(a)--(b).
Figure\,\ref{fig:evolution}(a) displays the variation of the blob number with time.
It can be seen that at least five blobs appear in the current sheet from 12:53 UT to 13:08 UT around the flare peak time.
We also obtained the light curves of AIA 94\,\AA\ and 211\,\AA\ in the flare loop region just below the current sheet
  (see the black contour in Figure\,\ref{fig:evolution}(d)) and showed them in Figure\,\ref{fig:evolution}(c).
The blob number peaks about five minutes after the peak time of the GOES X-ray flux and the AIA 94\,\AA\ light curve,
  and suddenly decreases afterwards.
However, the AIA 211\,\AA\ light curve peaks close to the peak time of the blob number.
Figure\,\ref{fig:evolution}(b) displays the temporal variations of the blob width and velocity.
The blob width seems to have little variation.
On the other hand, the blob velocity varies a lot and is up to 350\,\kms\ at the early stage of the eruption.
The error bars are derived from the standard deviations.

We also analyzed the EUV intensity variation along the current sheet.
Using a fast Fourier transform technique, we obtained the one-dimensional spectrum distributions of the AIA 304\,\AA, 171\,\AA, 193\,\AA, and 211\,\AA\ intensity along the current sheet.
The spectrum distributions of AIA 211\,\AA\ and 171\,\AA, as shown in Figure\,\ref{fig:powerindex},
  both show a power-law behavior and a decrease of the fitting spectral index \al\ with the increase of the blob number.
The spectral index \al\ is fitted in the wavenumber range of 0.2--1.67, i.e., the inverse of the blob width (0.6--5.0 Mm).
This value of \al\ is similar to that reported by \cite{2018ApJ...866...64C} for an off-limb current sheet.
Around 12:53:57 UT (t1, the early stage of the current sheet),
  the spectrum distribution is steep with a spectral index \al\ of 1.56 (1.93) in AIA 211\,\AA\ (171\,\AA).
This time corresponds to the appearance of the plasma blobs.
During the evolution of the current sheet from 12:55:45 UT (t2) to 13:01:33 UT (t3),
  the spectrum distributions become flatter with a spectral index \al\ of 0.92--0.99 (1.11--1.21) in AIA 211\,\AA\ (171\,\AA), corresponding to the growth and the maximum of the blob number.
After the peak instance of the blob number, the spectral index \al\ is steady around 1.0, e.g., 0.99 (0.95) in AIA 211\,\AA\ (171\,\AA) at 13:03:09 UT (t4),
  which is similar to that of the AIA 171\,\AA\ synthetic intensity from a numerical model conducted by \cite{2020ApJ...897...64Y}.
The numericial studies of \cite{2019MNRAS.482..588Y} and \cite{2020ApJ...897...64Y} also showed a decrease of the spectral index when multiple small-scale structures appear in the current sheets.
In addition, the AIA 304\,\AA\ and 193\,\AA\ intensities along the current sheet also reveal a power-law behavior (not shown here).

\section{Discussion}
\label{sec:dis}

Previous studies of current sheets during solar eruptions mostly focus on limb observations in white-light and EUV passbands
  \citep[e.g.,][]{2011ApJ...727L..52R,2012ApJ...754...13S,2013MNRAS.434.1309L,2018ApJ...853L..18Y,2018ApJ...866...64C,2020A&A...644A.158P}.
Sporadic plasma blobs have been found in the post-CME current sheets \citep[e.g.,][]{2003ApJ...594.1068K,2005ApJ...622.1251L,2010ApJ...723L..28L,2012SoPh..276..261S,2013ApJ...771L..14G,2014ApJ...784...91L,2016ApJ...831...47S,2017ApJ...841...49C,2020ApJ...892..129L}.
Here, we report a case study that reveals detailed evolution of many fine structures in the current sheet around the peak of an eruptive flare.
In the EUV images, during about 24 minutes, at least 108 compact plasma blobs propagate bidirectionally from the reconnection site, which rises with a velocity of $\leqslant$69\,\kms.
In addition, we investigated the temporal variation of the blob number around the flare peak time,
  which has not been provided by the previous studies.

In general, direct observation of fine structures in a current sheet is very difficult due to the limited resolution of current instruments.
In this study, we have identified more than one hundred compact plasma blobs with an average width of $\sim$1.7 Mm
  in a current sheet of a solar eruption from EUV observations.
Previous AIA observations typically revealed only several plasmoids or blobs in current sheets of solar eruptions
  \citep{2016ApJ...821L..29Z,2018ApJ...869..118D,2019SciA....5.7004G}.
\cite{2020A&A...644A.158P} identified 20 sunward and 16 anti-sunward plasmoids using the enhanced AIA 131\,\AA\ images,
  but in about two hours during the post-impulsive phase of the solar eruption.
\cite{2019ApJ...887L..34F} also investigated the current sheet of this solar eruption using spectropolarimetric data.
They found that small-scale magnetic field structures on scales of $\leqslant$6 Mm could be created by plasmoid formation during the reconnection process.
These plasmoids are more diffuse and larger than the blobs identified in our study.
With white-light observations, plasma blobs have also been identified in current sheets of solar eruptions
  \citep[e.g.,][]{2013ApJ...771L..14G,2020ApJ...892..129L,2020A&A...644A.158P},
  which are far away from the reconnection site as diffuse structures and two orders of magnitude larger than ours.

From the AIA 211\,\AA\ images, we have measured the physical parameters for the plasma blobs.
The blob width is in the range of 0.8--3.0 Mm with an average value of 1.7 Mm,
  which is similar to those reported by \cite{2018ApJ...869..118D} and \cite{2019SciA....5.7004G}, and 
  smaller than those reported by \cite{2016ApJ...821L..29Z} and \cite{2020A&A...644A.158P}.
In contrast, from off-limb observations, the widths of the current sheets of solar eruptions generally range from 10 Mm to about 100 Mm at heights of 0.27--1.16 solar radii from the solar surface \citep[see ][]{2015SSRv..194..237L}.
The propagating velocity of the blobs is estimated to be 50--550\,\kms\ with an average value of 190\,\kms,
  which is consistent with that from the EUV observations \citep{2016ApJ...821L..29Z,2020A&A...644A.158P}.
In addition, the width of the plasma blobs identified in a current sheet of solar eruption seems to follow a power law distribution \citep{2013ApJ...771L..14G,2020A&A...644A.158P}
  and has a correlation between the width and velocity \citep{2020A&A...644A.158P}.
For the blob width in this study, however, we do not find a power law distribution or a correlation with other parameters,
  which might be due to that only small plasma blobs evolving for only several minutes in the low corona are identified.

What roles the fine structures within a current sheet play during the energy release process of solar eruptions is still an open question.
Numerical simulations of plasmoid instability have attempted to investigate the fine structures inside current sheets and their temporal behavior \citep[e.g.,][]{2015ApJ...799...79N,2019MNRAS.482..588Y}.
The observational features in this study appear to be consistent with some results of the numerical models 
  \citep{2019MNRAS.482..588Y,2020ApJ...897...64Y,2021ApJ...909...45Y,2022RAA....22h5010Z,2023ApJ...955...88Y}.
In the numerical studies conducted by \cite{2019MNRAS.482..588Y} and \cite{2023ApJ...955...88Y},
  the sunward flows and the the anti-sunward ones only appear below and above the principal X-points, respectively.
This agrees well with our observational feature in the time-distance diagram and the animation.
In addition, we found that the temporal variation of the blob number is similar to that of the AIA 211\,\AA\ light curve obtained from the flare loop region during the solar eruption.
The result from the power spectrum analysis in our study shows a similarity to that of the numerical studies by \cite{2019MNRAS.482..588Y,2020ApJ...897...64Y}.
The blob size is also similar to that in the numerical model conducted by \cite{2022RAA....22h5010Z}.
Our observational results suggest that plasmoid instability may play a key role in the energy release processes of this solar eruption.

The bidirectional flows or plasmoids have also been identified at one side of the principle X-point in other numerical studies \citep[e.g.,][]{2015ApJ...799...79N,2015ApJ...812...92N}.
The discrepancy between our observational and the latter numerical results might be explained as the following.
First, above the principal X-point, the rising current sheet may compensate for possible sunward flows/blobs.
Second, the current sheet is a three-dimensional structure. 
The LOS superposition effect and the projection effect may impact the apparent motions of the outflows within the current sheet in the plane of sky.

\section{Summary}
\label{sec:sum}

In this study, we have reported an on-disk observation of numerous fine structures in a current sheet around the peak time of an eruptive flare.
The current sheet appears in the low corona connecting the flare loops and the rising filament,
  and involves at least 108 compact plasma blobs propagating in both sunward and anti-sunward directions near the reconnection site.
The width, duration, projected velocity, temperature, and density of these blobs are $\sim$1.7$\pm$0.5\,Mm, $\sim$79$\pm$57\,s, $\sim$191$\pm$81\,\kms, $\sim$10$^{6.4\pm0.1}$ K, and $\sim$10$^{10.1\pm0.3}$ cm$^{-3}$, respectively.
Using the velocity ratio of inflows and outflows associated with the reconnection process, the reconnection rate is estimated to be 0.27--0.33.
In the meantime, the reconnection site rises with a velocity of $\leqslant$69\,\kms.
We have also analyzed the change of the blob number with time during the flare peak time,
  which varies in a way similar to that of the AIA 211\,\AA\ light curve obtained from the flare loop region.
These observational results suggest that plasmoid instability plays an important role in the energy release process of solar eruptions.

\begin{acknowledgements}
This work was supported by Strategic Priority Research Program of the Chinese Academy of Sciences (Grant No. XDB0560000), National Key R\&D Program of China No.\,2022YFF0503800, NSFC grant 12303057, China Postdoctoral Science Foundation No.\,2021M700246, and the New Cornerstone Science Foundation through the XPLORER PRIZE.
M.M. acknowledges DFG grants WI 3211/8-1 and WI 3211/8-2, project number 452856778, and was partially supported by the Bulgarian National Science Fund, grant No KP-06-N44/2.
This research is partially supported by the Bulgarian National Science Fund, grant No KP-06-N44/2.
H. C. Chen is supported by NSFC grants 12103005.
Zhentong Li is supported by the Prominent Postdoctoral Project of Jiangsu Province.
The authors thank Jing Ye, Pengfei Chen, Xiaoli Yan, Xin Cheng, Ruisheng Zheng, Zhenghua Huang for useful discussions.
AIA is an instrument onboard the Solar Dynamics Observatory, a mission for NASA's Living With a Star program.
SUTRI is a collaborative project conducted by the National Astronomical Observatories of CAS, Peking University, Tongji University, Xi'an Institute of Optics and Precision Mechanics of CAS and the Innovation Academy for Microsatellites of CAS.
ASO-S mission is supported by the Strategic Priority Research Program on Space Science, the Chinese Academy of Sciences, Grant No. XDA15320000.
SOHO is a mission of international cooperation between ESA and NASA.
\end{acknowledgements}

\bibliographystyle{aa}
\bibliography{bibliography}

\end{document}